\documentclass[fleqn,usenatbib]{mnras}

\usepackage{newtxtext,newtxmath}

\usepackage[T1]{fontenc}
\usepackage{multirow}
\DeclareRobustCommand{\VAN}[3]{#2}
\let\VANthebibliography\thebibliography
\def\thebibliography{\DeclareRobustCommand{\VAN}[3]{##3}\VANthebibliography}

\usepackage{graphicx}	
\usepackage{amsmath}	
\usepackage{xcolor}
\colorlet{RED}{red}

\title[]{A semi-analytical Ring-Arc model for Plutino perturbations}

\author[Chen \& Li]{
Yue Chen$^{1,2}$
and Jian Li$^{1,2}$\thanks{E-mail: ljian@nju.edu.cn}
\\
$^{1}$School of Astronomy and Space Science, Nanjing University, 163 Xianlin Avenue, Nanjing 210023, PR China\\
$^{2}$Key Laboratory of Modern Astronomy and Astrophysics in Ministry of Education, Nanjing University, Nanjing 210023, PR China
}

\date{Accepted XXX. Received YYY; in original form ZZZ}

\pubyear{2015}

\begin{document}
\label{firstpage}
\pagerange{\pageref{firstpage}--\pageref{lastpage}}
\maketitle

\begin{abstract}
Neptune's 2:3 mean motion resonance induces an asymmetric spatial distribution of Plutinos. We previously proposed a 9000-point‑mass arc model that successfully captures the resulting influence on planetary ephemerides. However, integrating thousands of point masses involves significant computational cost. In this paper, to more efficiently simulate the total perturbation of Plutinos, we develop a semi-analytical Ring-Arc model described by gravitational potentials. This model consists of an outer ring and two inner arcs, representing the resonant spatial distribution of Plutinos. The model parameters are derived from observed Plutinos and theoretical sampling, yielding consistent structures. Numerical simulations show that the perturbations on the Sun-Neptune and Sun-Saturn distances from the semi-analytical Ring-Arc model agree with those from the 9000-point-mass arc model within $4\%$ over 50 years, while outperforming a simple ring model in comparison with the bias-corrected Plutino population. Our semi-analytical approach preserves the dynamical characteristics while significantly enhancing computational efficiency, offering a state-of-the-art tool for ephemeris modelling.
\end{abstract}




\section{Introduction}
The precision of planetary ephemerides is critical for space exploration missions, which requires ongoing refinement of dynamical models \citep[e.g.][]{newhall1983102, pitjeva2001modern, fienga2008inpop06, pitjeva2014development, folkner2014planetary, fienga2020inpop}. Conventional ephemeris computations incorporate mutual perturbations of the Sun, the eight planets and their natural satellites, the Pluto-Charon system, along with perturbations arising from the solar oblateness $J_2$, the solar pressure, figure and tide effects, relativistic corrections, lunar librations, and so on \citep{pitjeva2014development}. In recent years, modern planetary ephemerides have been updated by incorporating gravitational perturbations of a large number of asteroids observed in the Solar system \citep{folkner2014planetary, fienga2008inpop06, pitjeva2014development, Tian2023}. Contributions mainly come from three asteroid populations: main belt asteroids (MBAs), Jupiter Trojans (JTs), and Kuiper belt objects (KBOs). MBAs are commonly modelled using the `Bigs + ring' approach. 
In this approach, the most influential MBAs (the `Bigs') are individually integrated, while the remaining numerous bodies are represented by a homogeneous ring  implemented as a gravitational potential \citep{Kuchynka2010, pitjeva2018masses, fienga2020inpop}. It must be noted that the validity of this ring approximation relies on the fact that the distribution of the MBAs exhibits gaps at locations corresponding to mean motion resonances (MMRs) with Jupiter. In contrast, asteroids elsewhere may accumulate in the MMRs with a planet and thus exhibit special spatial structures. For instance, a substantial population of Jupiter Trojans resides in the 1:1 MMR with Jupiter, occupying the L4 and L5 points with an asymmetric distribution. Based on this characteristic, \citet{li2018constructing} estimated that the total gravitational perturbation of Jupiter Trojans on the Earth–Mars distance reaches $\sim70$ m over a century.

In the Kuiper belt, a significant number of asteroids are trapped in Neptune's MMRs. In the previous work \citep{chen2024}, we began to examine whether the asymmetric spatial distribution induced by the MMRs with Neptune has a sufficient influence on planetary positions to be considered in ephemerides. 
Plutinos, which reside in Neptune's 2:3 MMR at the semimajor axis $a \sim 39.4$ AU, are considered in the first place, as they constitute the most populous resonant population with Neptune. A discrete point-mass arc model was employed to represent the Plutinos' resonant, non-uniform distribution. Compared with a homogeneous ring model, which was used in \citet{pitjeva2018mass}, a significant difference in planetary perturbations was revealed, especially when simulating high-eccentricity Plutinos. Taking the perturbation of observed Plutinos as the standard, we find that the difference in the Sun-Saturn distance for the point-mass arc model is on the order of 1 km, while for the ring model it is about twice as large. 
Based on \textit{Cassini} tracking data collected between 2004 and 2017, the accuracy of Saturn' positions has been significantly improved to about 0.025 km \citep{Park_2021,fienga2021}. Therefore, the $\sim$1 km difference between the two models already exceeds the current observational threshold for Saturn ($\sim 0.025$ km), making the point-mass arc model a practically necessary refinement rather than merely a theoretical expectation. 
However, the point-mass arc model, being discrete and comprising 9,000 point masses, can significantly slow down numerical integration in ephemeris calculations.
To address this inefficiency, we aim to develop a continuous model expressed via a gravitational potential to replace it. This approach maintains the characteristics of the resonant distribution while improving computational efficiency.

The rest of this paper is organised as follows. 
Sect. \ref{sec:model} develops a semi-analytical Ring-Arc model described by the gravitational potential that incorporates the resonant spatial distribution of Plutinos.
In Sect. \ref{sec:compare}, this new semi-analytical Ring-Arc model is evaluated by comparisons with the previous ring model, the point-mass arc model, and the bias-corrected observational data from currently known Plutinos. Finally, the conclusions are presented in Sect. \ref{sec:conclusion}.

\section{The semi-analytical Ring-Arc model}
\label{sec:model}
As mentioned in the Introduction, we aim to develop an alternative model based on evaluating the gravitational potential rather than integrating thousands of point masses to reduce the integration time. We will first analyse the spatial distribution of the Plutinos and then derive their total gravitational potential.

\subsection{Spatial distribution of Plutinos}
\label{sec:2.1}

\begin{figure}
    \centering
    \includegraphics[width=0.8\linewidth]{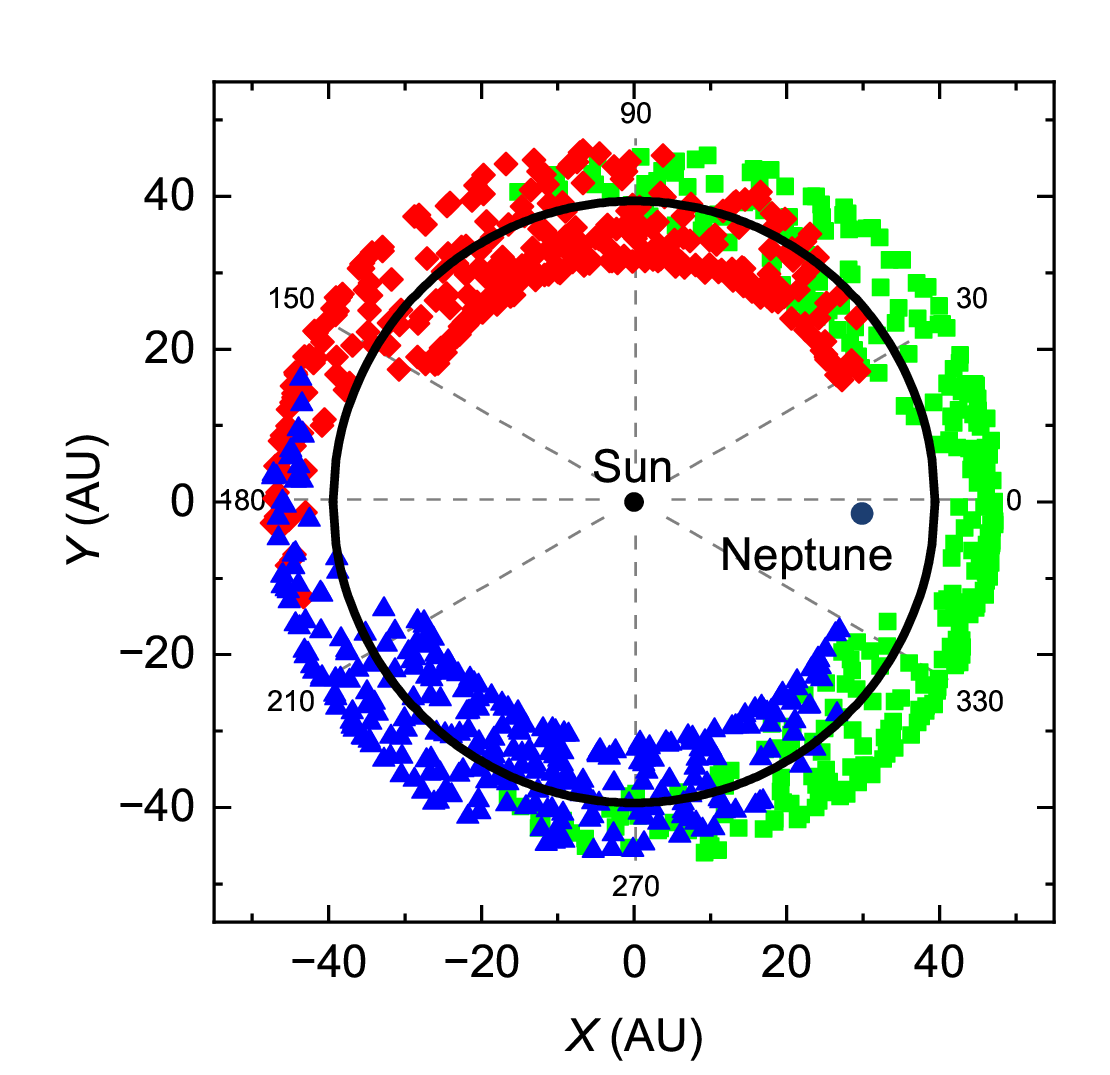}
    \caption{The simplified model for the spatial distribution of Plutinos. The parameter of the coplanar bodies are $a\in(39.1,39.7)$ AU, $e\in(0.1,0.3)$ and $\sigma\in$ [$60^{\circ}$, $300^{\circ}$] (green squares), $\sigma\in$ [$420^{\circ}$, $660^{\circ}$] (red diamonds), $\sigma\in$ [$-300^{\circ}$, $-60^{\circ}$] (blue up-triangles). The black circle indicates the heliocentric distance of $r = 39.4$ AU.}
    \label{fig:1}
\end{figure}

Here, we present a population of coplanar test Plutinos in Fig. \ref{fig:1}. The semimajor axes of these bodies are randomly sampled from a uniform distribution between $39.1$ and $39.7$ AU. 
The eccentricities of the test Plutinos are selected from a uniform distribution between $0.1$ and $0.3$. As shown in Fig. 8 of our previous work \citep{chen2024}, more than 90\% of the observed Plutinos have eccentricities within this range. The eccentricity primarily determines the radial extent of the arcs: smaller eccentricities result in narrower radial distributions, whereas larger eccentricities lead to wider ones. We note that the eccentricity range determines the overall radial structure and provides the initial basis for designing the geometric configuration of the semi-analytical Ring-Arc model. Nevertheless, the density distribution within the arc along the radial direction (i.e. the inhomogeneous eccentricity distribution) should also be considered. In the following sections, when determining the parameters of the semi-analytical Ring-Arc model, we further take into account the effects of different eccentricity distributions.

In addition, we need to determine the resonant angle $\sigma$, which is defined as 
\begin{equation}
 \sigma=3\lambda-2\lambda_N-\varpi   
\end{equation}
where $\lambda$ denotes the mean longitude and the subscript $N$ refers to Neptune, and $\varpi$ is the longitude of the perihelion of the test Plutino. To ensure that the test Plutinos reside stably in Neptune's 2:3 MMR, we assume that the resonant angle $\sigma$, which librates about the resonant centre $\sigma_0=180^\circ$, has a resonant amplitude $A$ uniformly distributed between 0 and $120^\circ$. The upper limit of $A$ is adopted to $120^\circ$ considering the stability of Plutinos over the age of the Solar system \citep{Levision1995,Nesvorny2000,li2014study,Ito2025}. Consequently, the resonant angle is restricted to $\sigma\in[60^\circ,300^\circ]$. Once $\sigma$ is specified, with $\varpi$ randomly drawn from 0 to $360^\circ$, the mean anomaly $M$ of the test Plutino is computed by
\begin{equation}
    M=\frac{1}{3}(\sigma +2\lambda_N-2\varpi).
\end{equation}

It is essential to note that when the phase space of the 2:3 resonance is unfolded, three resonant islands emerge, corresponding to $\sigma\in$ [0, $360^{\circ}$], [$-360^{\circ}$, 0], and [$360^{\circ}$, $720^{\circ}$], respectively \citep{li2023study}. Therefore, in addition to the commonly considered range of $[0, 360^\circ]$, $\sigma$ is allowed to vary within the other two ranges by associating the angle $\sigma=\sigma\pm360^\circ$. Although such displacements in $\sigma$ preserve the resonant behaviours of Plutinos, the mean longitude $M$ would change significantly. As depicted in Fig. \ref{fig:1}, different coloured symbols are used to distinguish between the various ranges of $\sigma$.

This simplified model demonstrates a distinct spatial distribution of Plutinos. 
Given that the resonant angle $\sigma$ for Plutinos oscillates around $\sigma_0 = 180^\circ$, these objects tend to occupy longitudes displaced by $\pm 90^\circ$ relative to Neptune when they reach perihelion \citep{jewitt1998large}. Based on this, \citet{Chiang2002} introduced simplified toy models to describe the spatial distribution of Plutinos. 
By incorporating the libration amplitudes of $\sigma$, they demonstrated that Plutinos are distributed over a range of longitudes, peaking at $\pm 90^\circ$ from Neptune, and that the extent of the most concentrated population (shown as the red and blue arcs within the black circle in Fig. \ref{fig:1}) is directly correlated with the resonant amplitude $A$.
By adopting the nominal location of Neptune's 2:3 MMR, specifically at heliocentric distance $r = 39.4$ AU, as a dividing boundary, the spatial distribution of Plutinos can be characterised as illustrated in Fig. \ref{fig:1}: within the boundary (black circle), they are modelled as two arcs centred at longitudes displaced by $\pm 90^\circ$ relative to Neptune, while beyond the boundary they are mimicked by a complete ring. Accordingly, we can derive a semi-analytical model to represent the gravitational perturbations exerted by Plutinos on a planet.

\subsection{Modelling Plutinos and parameter selection}
\label{sec:2.2}

\begin{table*}
    \centering
    \begin{tabular}{c|c|c|c|c|c|c}
        \hline
       Ring-Arc model  &  $r_{ring}$ (AU) & $r_{arc}$ (AU) & $m_{ring}$~:~$m_{arc}^1$~:~$m_{arc}^2$ & $\phi_{ring}$ & $\phi_{arc}^1$  & $\phi_{arc}^2$ \\
        \hline
        DIT                 & 44.82 & 34.86 & $2.78:1:1$ & $[0^\circ, 360^\circ]$ & $[30^\circ, 150^\circ]$ & $[210^\circ, 300^\circ]$ \\
        DIS($e=0.1-0.3$) & 44.35 & 34.66 & $2.78:1:1$ & $[0^\circ, 360^\circ]$ & $[30^\circ, 150^\circ]$ & $[210^\circ, 300^\circ]$\\
        DIS($e=0.2$)         & 44.38 & 34.59 & $2.78:1:1$ & $[0^\circ, 360^\circ]$ & $[30^\circ, 150^\circ]$ & $[210^\circ, 300^\circ]$\\
        \hline
    \end{tabular}
    \caption{Parameters for the semi-analytical Ring-Arc model. The superscripts ``1'' and ``2'' of $m_{arc}$ and $\phi_{arc}$ refer to the arcs leading and trailing Neptune, respectively. For $\phi_{ring}$, $\phi_{arc}^1$, and $\phi_{arc}^2$, the left and right endpoints of the interval correspond to the integration limits $\phi_0$ and $\phi_1$ in Eq (\ref{eq:ring}), respectively.}
    \label{tab:1}
\end{table*}

Based on the characteristics of the spatial distribution of Plutinos analysed above, we developed a continuous model combining a ring and two arcs, called the semi-analytical Ring-Arc model. We modelled the two continuous arcs in the region inside $r = 39.4$ AU, and a continuous ring in the outer region (see Fig. \ref{fig:1}). 
The test particles shown in Fig. \ref{fig:1} are used only to illustrate the spatial distribution of the Plutino population and to motivate the geometry of our semi-analytical Ring-Arc model. They will neither be employed in the subsequent numerical simulations nor shown in the associated figures.
For a continuous arc or a continuous ring, the corresponding gravitational potential at the location $(x',y', z')$ can be described as
\begin{equation}
  \begin{aligned}
	&U(x',y',z',t)=-\int_r\int_{\phi_0}^{\phi_1}\frac{G\rho r}{D}drd\phi, \\
	&D=\sqrt{(rcos\phi-x')^2+(rsin\phi-y')^2+z'^2},
    \label{eq:ring}
  \end{aligned}
\end{equation}
where $G$ is the gravitational constant, $\rho$ is the linear density, and $(r,\phi)$ are the polar coordinates of the arc/ring with respect to the Sun. Notably, the two parameters $\rho$ and $r$ will be treated as fixed, resulting in Eq. (\ref{eq:ring}) with only one variable $\phi$. Consequently, the integration reduces to a one-dimensional form and can be performed using the Romberg algorithm.

For the continuous ring beyond $r = 39.4$ AU, the lower and upper limits of integration for the angular coordinate $\phi$ in Eq. (\ref{eq:ring}) are $\phi_0=0$ and $\phi_1=360^\circ$, respectively. For the inner continuous arcs, their lengths are determined by the resonance amplitudes $A$ of the test Plutinos. This follows from the physical meaning of the resonant angle $\sigma$: when a Plutino is at its perihelion (i.e. $M=0$), we have
\begin{equation}
\sigma=2\lambda-2\lambda_N+(\lambda-\varpi)=2(\lambda-\lambda_N)+M=2(\lambda-\lambda_N).
\end{equation}
Thus, $\sigma$ measures the longitude separation of the Plutino from Neptune. When $\sigma$ librates within the range $[\sigma_0-A,\sigma_0+A]$, given $\sigma_0=180^\circ$ and $A\le120^\circ$, its maximum variation corresponds to the longitude separation of $30^\circ$ to $150^\circ$. This configuration is known as the phase-protection mechanism. As shown in Fig. \ref{fig:1}, if the Plutinos are at perihelion (i.e within $r = 39.4$ AU), their angular coordinates satisfy $\phi\in[30^\circ, 150^\circ]$, indicated by the red arc inside the black boundary. Considering the two additional resonant configurations with $\sigma\pm360^\circ$, this gives rise to the blue arc, also inside the black boundary, with $\phi\in[210^\circ, 300^\circ]$. Accordingly, these ranges determine the lower and upper limits, $\phi_0$ and $\phi_1$, of the two continuous arcs in our model.

Subsequently, we estimate the linear density $\rho$ and radius $r$ in Eq. (\ref{eq:ring}) by treating the outer ring and inner arcs separately, with parameters $\rho_{ring}$, $r_{ring}$ and $\rho_{arc}$, $r_{arc}$, respectively. Similarly to our previous work \citep{chen2024}, we first identify the Plutino population from the observed KBOs catalogued in the Minor Planet Center (MPC) database\footnote{https://minorplanetcenter.net/iau/lists/TNOs.html}. Through 1 Myr orbital integration within the framework of the outer Solar system, 460 objects are identified as Plutinos, characterised by the libration of the resonant angle $\sigma$. However, this number is significantly smaller than the theoretical prediction \citep{alexandersen2016carefully}, and the distribution of Plutinos may be biased because they are more likely to be detected near the perihelion. Secondly, to mitigate this possible impact, we enlarge the sample size using two methods developed in \citet{chen2024}:

(1) Disperse In Time (DIT): 
The Birkhoff ergodic theorem \citep{birkhoff1931} provides the theoretical motivation: for an ergodic system, time averages converge to ensemble averages. Since our integration time is much longer than the secular precession period of the orbits of Plutinos, each orbit has sampled a large fraction of its accessible phase space -- a practical approximation to ergodic behaviour. Accordingly, the influence of the observational bias in the currently observed Plutino sample could be mitigated by taking a time average. 
However, this method may not completely eliminate the bias, as the extent of mitigation depends on the type of observational bias inherent in the initial conditions. For example, Plutinos with smaller perihelion distances are more likely to be detected, as noted by \citet{chen2024}, and the 2:3 MMR can cause a skew in their distribution along the longitude. Consequently, observational bias may persist to some extent even after long-term ensemble averaging.

To implement the DIT method, for the sake of simplicity, we numerically approximate the time average rather than computing it exactly.
We first integrate the 460 observed Plutinos over 11 Myr and randomly select 10 epochs between 10 Myr and 11 Myr. 
For each selected epoch, we record the Plutinos’ positions relative to Neptune, and then combine all these positions over the 10 epochs to form the overall distribution. This multi-epoch sampling effectively mimics the long-term time average of the population, thereby reducing the observational bias inherent in the current snapshot of observed Plutinos. Then, under this ``unbiased'' distribution, we perform a Gaussian fit to derive $r_{ring}$ using the Plutinos in the outer part beyond the $r=39.4$ AU boundary, while $r_{arc}$ is derived from those in the inner part. In this way, the mean heliocentric distances are obtained as $r_{ring}=44.82$ AU and $r_{arc}=34.86$ AU. 
In addition, the number ratio of Plutinos in the outer region to those in the inner region is $1:0.72$. Assuming that each Plutino has the same mass, this corresponds to a mass ratio of $m_{ring}:m_{arc}^1:m_{arc}^2=2.78:1:1$ between the ring and the leading and trailing arcs, respectively. We note that this ratio is a geometrical counting result rather than a dynamically derived parameter. Specifically, it is obtained by simply counting the sampled particles in the outer region (beyond $r=39.4$ AU) and the two inner arcs.

(2)  Disperse In Space (DIS):
We first generate a group of planar test Plutinos with semimajor axes $a=39.4$ AU, eccentricities $e\in (0.1,0.3)$, and resonance amplitudes $A \in [0,120^\circ]$. The orbital parameters $\varpi$ and $M$ are selected following the procedure described in Sect. \ref{sec:2.1}. The test Plutinos are then divided into outer-ring and inner-arc populations relative to the $r=39.4$ AU boundary. The mean heliocentric distances of these two populations yield $r_{ring}=44.35$ AU and $r_{arc}=34.66$ AU. This case of the Ring-Arc model is denoted as ``DIS($e=0.1-0.3$)''. Similarly, we generate another group of test Plutinos with the same parameters, but their eccentricities are fixed at $e=0.2$. This case, denoted as ``DIS($e=0.2$)'', results in mean heliocentric distances $r_{ring}=44.38$ AU and $r_{arc}=34.59$ AU for the ring and arcs, respectively. 
Finally, we find it remarkable that, for both DIS cases, the mass ratio between the ring and the leading and trailing arcs is also $2.78:1:1$, exactly the same as that obtained above for the DIT case. 
This suggests that it may reflect a robust geometric feature of the Plutino spatial distribution.

The parameters of the semi-analytical Ring-Arc model derived above are summarised in Table \ref{tab:1}. 
The last three columns specify the range of the angular coordinate $\phi$ associated with the ring and the two arcs, thereby defining the integration limits $\phi_0$ and $\phi_1$ in Eq (\ref{eq:ring}). Then, to calculate the linear density $\rho$, it remains to determine the total mass of the Plutinos, $M_{plu}$. Following \citet{chen2024}, we adopt $M_{plu}=0.01M_\oplus$, which corresponds to approximately $1/6$ of the total mass of the KBOs \citep{di2020analysis}. Based on the mass ratios and $\phi$ ranges listed in Table \ref{tab:1}, we obtain the densities $\rho_{ring}=5.8\times 10^{-3} M_\oplus/(2\pi r)$ and $\rho_{arc}^1=\rho_{arc}^2=2.1\times 10^{-3} M_\oplus/(2\pi r/3 )$. 
We briefly note that, although the precise mass value $M_{plu}$ remains uncertain, our previous work \citep[][Sect. 3.3]{chen2024} demonstrated that the perturbation of the Plutino population on the Sun-Neptune distance scales nearly linearly with $M_{plu}$, which is also supported by \citet{Kuchynka2010} and \citet{li2018constructing}. Since the linear scaling holds, the resulting perturbation effects can easily be extrapolated.

\section{Comparison with point-mass models and observed Plutinos}
\label{sec:compare}
As in \citet{chen2024}, we quantify the gravitational effect of Plutinos on Neptune, the planet closest to them. Specifically, we evaluated the perturbation by comparing the Sun–Neptune distance obtained with and without the gravitational influence of the Plutinos. We define the difference as
\begin{equation}
\Delta d_{SN}=d_{SN1}-d_{SN0},
\label{dsn}
\end{equation}
where $d_{SN1}$ and $d_{SN0}$ represent the Sun–Neptune distances from simulations including and excluding the perturbations from the Plutinos, respectively. Thus, $\Delta d_{SN}$ directly measures the change in the Sun–Neptune distance induced by the Plutino population.
In addition, since the accuracy of Saturn’s ephemeris has been significantly improved by the in situ measurements from \textit{Cassini}, we also consider the corresponding change in the Sun–Saturn distance, defined as 
\begin{equation}
\Delta d_{SS}=d_{SS1}-d_{SS0},
 \label{dss}  
\end{equation}
where $d_{SS1}$ and $d_{SS0}$ denote the Sun-Saturn distances calculated with and without perturbations from the Plutinos, respectively.

The dynamical model adopted in our calculations consists of the Sun and the four Jovian planets, with the optional inclusion of the Plutinos. The contribution of terrestrial planets to this model is small enough to be neglected; further details can be found in our previous work \citep{chen2024}. For numerical calculations, we employ the 19th-order Cowell prediction-correction algorithm (PECE) with a time-step of 10 days, chosen according to the orbital period of the innermost body in the model (i.e. Jupiter) \citep{tian1993adams, li2018constructing}. Within this $N$-body framework, we take into account gravitational interactions among the Sun, planets, and Plutinos,  whereas the self-gravity of the Plutinos is neglected.

\subsection{Comparison with point-mass arc model}

\begin{figure*}
    \centering
    \includegraphics[width=\linewidth]{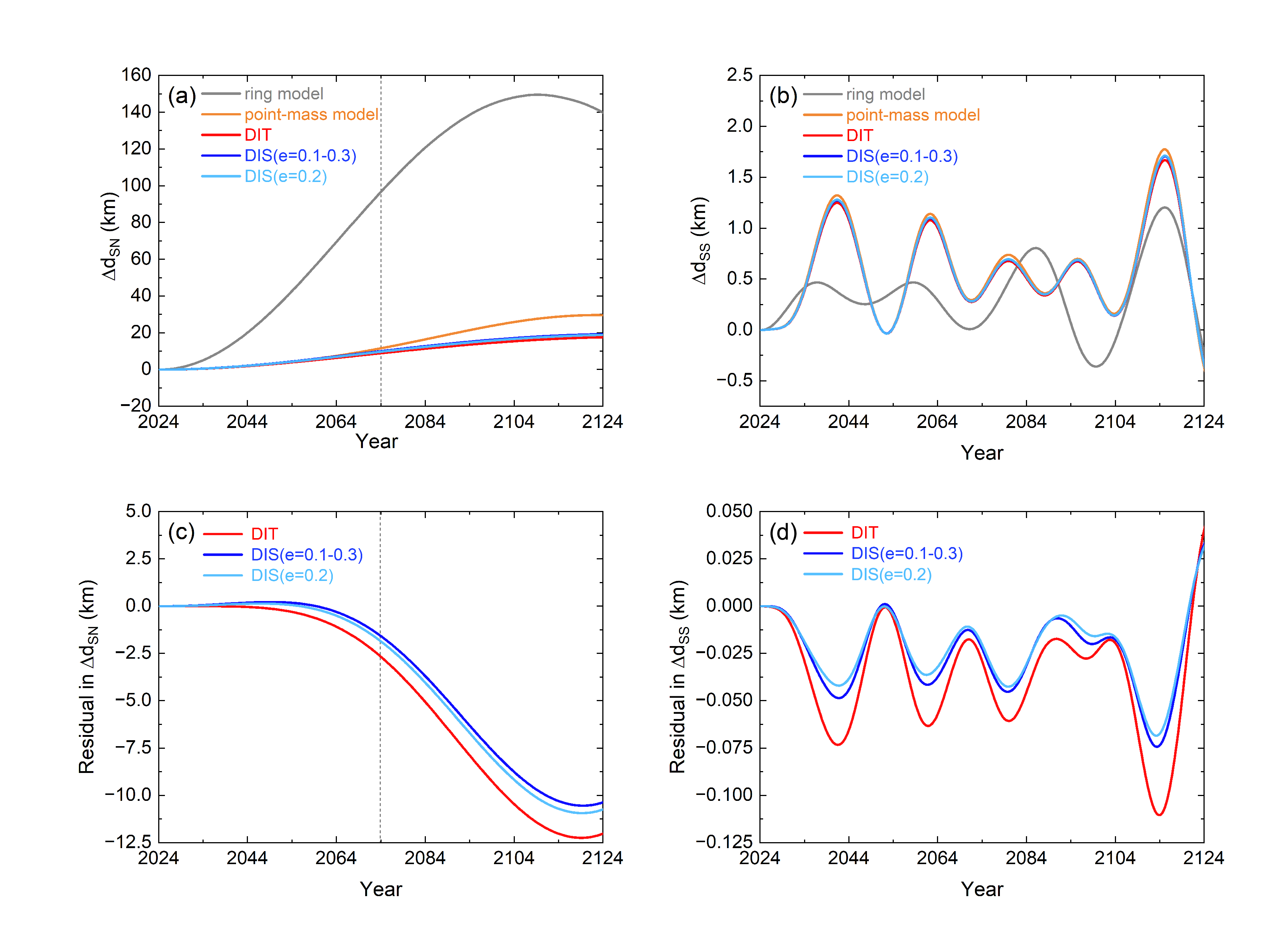}
    \caption{Perturbations on the Sun-Neptune distance (panel a) and the Sun-Saturn distance (panel b) induced by the ring model (grey), the point-mass arc model (orange), and the semi-analytical Ring-Arc models denoted by DIT (red), DIS($e=0.1-0.3$) (dark blue), and DIS$(e=0.2)$ (light blue). Relative to the point-mass benchmark indicated by the orange curve in panels (a) and (b), panels (c) and (d) show the corresponding differences between the DIT/DIS models and the benchmark.}
    \label{fig:model}
\end{figure*}

We first present the changes in planetary positions induced by the semi-analytical Ring-Arc model under different parameter choices, and then compare the results with those obtained from the ring model and the point-mass arc model. In the conventional ring model, the Plutino population is represented as a one-dimensional continuous heliocentric ring, whose gravitational potential is described by Eq. (\ref{eq:ring}) with the parameters $\phi_0=0$, $\phi_1=2\pi$, $r=39.4$ AU, and $\rho=M_{plu}/(2\pi r)$. To improve upon the limitations of the ring model, \citet{chen2024} introduced a point-mass arc model, in which 9000 point masses share the Plutino orbits, with semimajor axes $a=39.4$ AU, eccentricities $e=0.2$, inclinations $i=0$, and resonant amplitudes $A\le120^{\circ}$. We note that both the ring model and the point-mass arc model are taken directly from \citet{chen2024}; they each adopt a total mass of $M_{plu}=0.01M_{\oplus}$, consistent with that used in the semi-analytical Ring–Arc model. 

For the three types of Plutino models described above, we calculated their perturbations on the Sun-Neptune and Sun-Saturn distances. In \citet{chen2024}, we demonstrated that the point‑mass arc model with the given parameters provides the most representative approximation of the real Plutinos. 
As shown in Fig. \ref{fig:model}(a) and (b) for the perturbations on the Sun-Neptune and Sun-Saturn distances, respectively, the point-mass arc model (orange curve) represents the closest approximation to the ``true'' solution. Compared with the point-mass arc benchmark, the ring model (grey curve) exhibits significant deviations and therefore does not provide an accurate representation, whereas the semi-analytical Ring-Arc models (DIT in red, DIS ($e=0.1-0.3$) in dark blue, and DIS ($e=0.2$) in light blue) all agree closely with the benchmark. To better visualise the differences between the orange (point-mass) curve and the DIT/DIS curves, panels (c) and (d) of Fig. \ref{fig:model} plot the individual residuals of the DIT/DIS models relative to the point-mass benchmark. Within the first 50 years, for the Sun–Neptune distance change $\Delta d_{SN}$ (Fig. \ref{fig:model}(c)), the maximum difference between the semi-analytical Ring-Arc models and the point-mass arc model is only 2.8 km, while for the Sun–Saturn distance change $\Delta d_{SS}$ (Fig. \ref{fig:model}(d)), the difference is less than 0.05 km, corresponding to a relative error of only $\sim4\%$.
These results indicate that the semi-analytical Ring‑Arc models accurately reproduce the 9000-point-mass arc model while substantially reducing computational cost. Interestingly, all three semi-analytical Ring-Arc models, despite having slightly different parameters, result in almost identical perturbations: the maximum differences are 1 km in $\Delta d_{SN}$ and 0.006 km in $\Delta d_{SS}$. Therefore, the parameters of the semi-analytical Ring-Arc models (see Table \ref{tab:1}), whether derived from observational (i.e. DIT) or theoretical (i.e. DIS) approaches, are acceptable. 
Since the three semi-analytical Ring-Arc models yield very similar results (see Fig. \ref{fig:model}), for the sake of clarity and brevity, the DIT model is adopted as a representative case for the subsequent study.

\subsection{Comparison with observed Plutinos}

\begin{figure}
\includegraphics[width=\columnwidth]{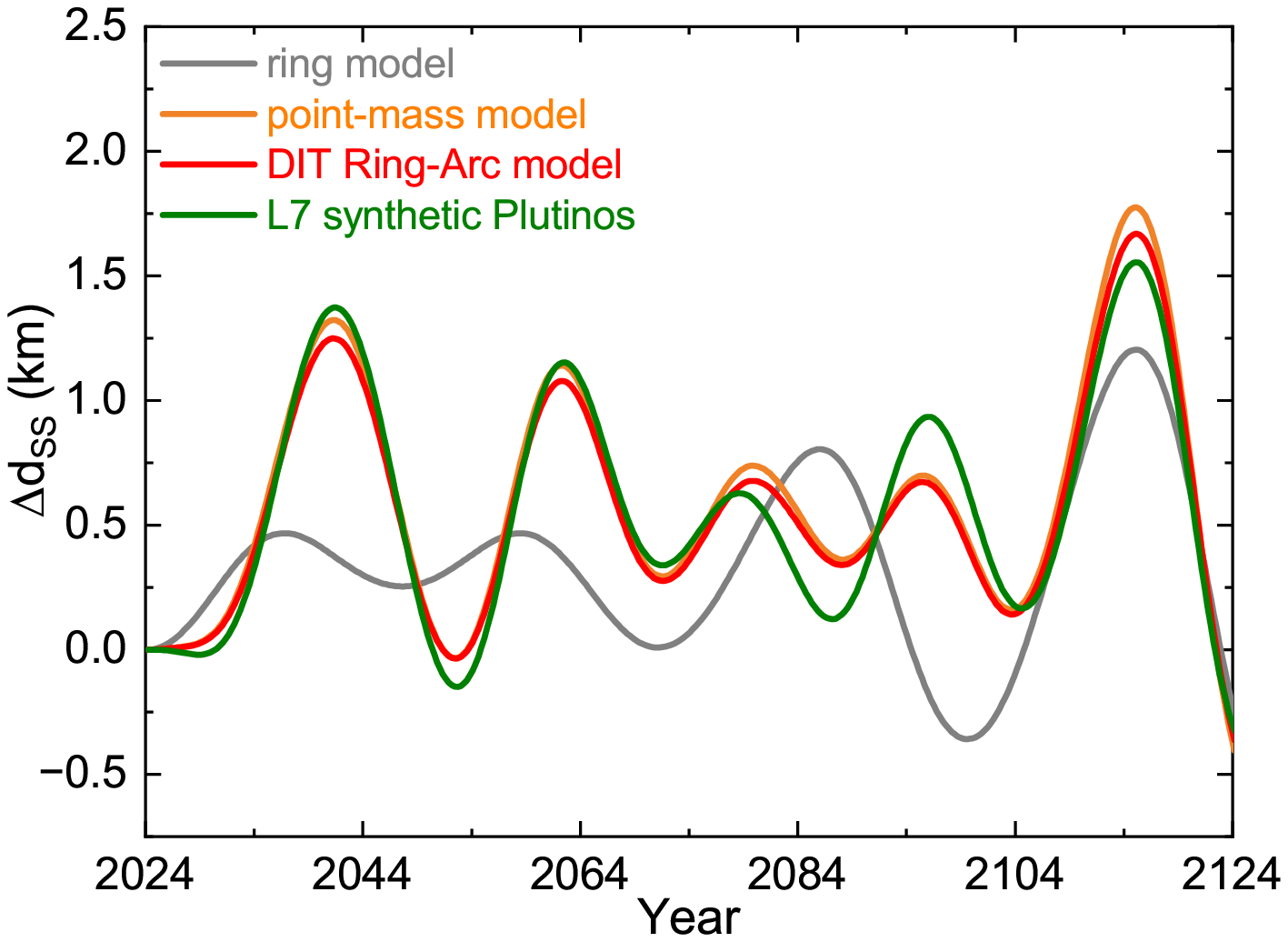}
\caption{Perturbations on the Sun-Saturn distance induced by the ring model (grey), the point-mass arc model (orange), the DIT Ring-Arc model (red), and the L7 synthetic Plutinos (green).
}
\label{fig:real}
\end{figure}

As mentioned in Sect. \ref{sec:2.2}, only 460 observed Plutinos have been identified, whereas the theoretically estimated population with diameters $D>100$ km exceeds 9000 \citep{alexandersen2016carefully}. This observational incompleteness may lead to a biased spatial distribution of the real Plutino population. 
To compensate for observational biases, the population of 2:3 resonant KBOs from the L7 synthetic model \citep{L7Kavelaars2009,L7Petit2011,L7Gladman2012} is used to represent the observed Plutinos. This model, based on the Canada-France Ecliptic Plane Survey (CFEPS), provides a calibrated population of resonant KBOs down to very small sizes, corresponding to an $H_g$ magnitude of 8.5. Since the L7 synthetic model corrects for observational biases by generating objects that have not yet been observed, it is expected to reflect the intrinsic orbital distribution of the Plutino population and is therefore adopted for comparison with our semi-analytical Ring-Arc model.
The use of the L7 synthetic model offers two main advantages. First, it enables a direct comparison between the debiased Plutino population at the present epoch and our semi-analytical Ring-Arc model. Second, the L7 Plutinos exhibit a substantial inclination distribution, whereas our model is coplanar. Consequently, when assessing the gravitational influence of the real Plutino population, the effects of its vertical distribution are implicitly taken into account.


Given the current low observational accuracy of Neptune's position, a direct comparison of the Sun–Neptune distance between theoretical models and observed Plutinos is of limited significance. By contrast, thanks to the high-precision measurements from the Cassini mission, the position of Saturn is much more precisely constrained. Therefore, it is preferable to focus on the change in the Sun-Saturn distance, $\Delta d_{SS}$, induced by different models and Plutino populations.

As shown in Fig. \ref{fig:real}, the $\Delta d_{SS}$ curves produced by the point-mass arc model (orange), the DIT Ring-Arc model (red), and the L7 synthetic model (green) show excellent agreement with each other, while they differ significantly from the ring model (grey). Within the first 50 years, the red curve from the DIT Ring-Arc model closely matches the orange curve from the point-mass arc model. Taking the green $\Delta d_{SS}$ curve from the L7 synthetic Plutinos as the benchmark, the maximum discrepancies of both the DIT Ring-Arc model and the point-mass arc model are less than 0.12 km, which is significantly smaller than the 1.02 km discrepancy of the ring model.
Given that the $\sim$1 km difference between the arc and ring models is about 40 times larger than the $\sim$0.025 km accuracy of Saturn's position measurements obtained from the \textit{Cassini} mission, such an improvement is well within the reach of current observational capabilities. This discrepancy confirms that resonance-induced asymmetries must be taken into account in modern planetary ephemerides.

Consequently, it is reasonable to conclude that the semi-analytical Ring-Arc model can replace the 9000-point-mass arc model developed in \citet{chen2024}. It not only provides a highly reliable representation of the perturbations caused by real Plutinos on planetary positions, but also, more importantly, requires significantly less computational cost for planetary ephemeris calculations. For the 50-year simulations performed above, the semi-analytical Ring-Arc model runs with an average CPU time of 0.96 seconds, compared to 125 seconds for the 9000-point-mass arc model, implying a speed‑up factor of about 130.

Finally, we note from Fig. \ref{fig:real} that, during the final period (around 2074–2104), the L7 synthetic Plutino model deviates from the point-mass model and the DIT Ring-Arc model. This deviation is likely caused by the accumulation of model errors over time. Therefore, our conclusion regarding the validity of the Ring-Arc model is primarily based on the first 50 years, a timescale sufficient for current high-precision ephemeris applications.

\section{Conclusions}
\label{sec:conclusion}

In this study, we have developed a continuous model, referred to as the semi-analytical Ring-Arc model, to efficiently simulate the perturbations induced by Plutinos on the positions of the Jovian planets. This model consists of an outer continuous ring and two inner continuous arcs, which are expressed as analytical gravitational potentials, and captures the intrinsic spatial distribution of Plutinos arising from their 2:3 MMR with Neptune.

We derived three sets of parameters for the semi-analytical Ring-Arc model based on different dispersion strategies: one from temporal sampling of the observed Plutinos (denoted DIT), and the other two from theoretical 2:3 resonant populations (denoted DIS). All approaches yield consistent structural parameters and mass ratios for the ring and arcs. Numerical simulations show that the perturbations on the Sun-Neptune and Sun-Saturn distances induced by the DIT and DIS semi-analytical Ring-Arc models closely match those from the discrete point-mass arc model constructed and validated in \citet{chen2024}, with relative errors remaining within $4\%$ over 50 years. Moreover, compared with the perturbation caused by the bias-corrected Plutino population from the L7 synthetic model, the semi-analytical Ring-Arc model significantly outperforms the conventional ring model in representing these effects.
It is noteworthy that, although the L7 Plutinos exhibit a realistic inclination distribution whereas our Ring-Arc model is coplanar, the latter still reproduces the gravitational perturbations of the Plutino population remarkably well. This demonstrates that neglecting the vertical distribution in the Ring-Arc model has only a minor effect on the perturbations considered here.

These results confirm that the semi-analytical Ring-Arc model provides an accurate representation of the gravitational influence of Plutinos. Furthermore, this model involves only the gravitational potential integral and is therefore significantly more computationally efficient than the previous point-mass arc model, which comprises as many as 9000 point masses. As a result, the semi-analytical Ring-Arc model offers a state-of-the-art approach for incorporating Plutinos into planetary ephemeris calculations.

\section*{Acknowledgements}
This work was supported by the National Natural Science Foundation of China (Nos. 12473061, 11973027, 12150009), and the China Manned Space Program (CMS-CSST-2025-A16). The authors would like to express their thanks to the anonymous referee for the valuable comments that helped considerably improve the manuscript.

\section*{Data Availability}
The data underlying this article are available in the article and in its online supplementary material.



\bibliographystyle{mnras}
\bibliography{example} 







\bsp	
\label{lastpage}
\end{document}